\documentclass{sig-alternate}
\usepackage{amsmath,amsfonts,bbm,subfigure}
\usepackage{algorithm,algorithmic,enumitem,comment}

\usepackage{tikz}
\usetikzlibrary{arrows}
\usepackage[htt]{hyphenat} 

\newtheorem{theorem}{Theorem}

\def\EE{{\mathbb{E}}}
\def\PP{{\mathbb{P}}}

\newcommand{\R}{{\mathbb R}}

\newcommand{\PR}{\pi}

\newcommand{\outneighbors}[1]{\mathcal{N}^{out}(#1)}
\newcommand{\inneighbors}[1]{\mathcal{N}^{in}(#1)}
\newcommand{\outdegree}[1]{d^{out}(#1)}
\newcommand{\indegree}[1]{d^{in}(#1)}
\newcommand{\epr}{r_{\text{max}}}
\newcommand{\rmax}{\epr}

\newcommand{\abs}[1]{\left \lvert #1 \right \rvert}
\newcommand{\size}[1]{\left \lvert #1 \right \rvert}
\newcommand{\pn}[1]{\left ( #1 \right )}
\newcommand{\bk}[1]{\left [ #1 \right ]}
\newcommand{\pfail}{p_{\text{fail}}}
\newcommand{\norm}[1]{\left\lVert #1 \right\rVert}

\newcommand{\pluseq}{\mathrel{+}=}
\newcommand{\john}{John}

\newcommand{\nsample}{n_{s}}

\newcommand{\bippr}{\texttt{Bi\-directional-PPR}}
\newcommand{\bipprgrouped}{\texttt{BiPPR-Precomp-Grouped}}
\newcommand{\bipprsampling}{\texttt{BiPPR-Precomp-Sampling}}
\newcommand{\bipprprecomp}{\texttt{BiPPR-Precomp}}
\newcommand{\bipprbasic}{\bippr{}}

\newcommand{\mc}{\texttt{Monte-Carlo}}
\newcommand{\rpush}{\texttt{Approx-Contri\-butions}}

\newfont{\mycrnotice}{ptmr8t at 7pt}
\newfont{\myconfname}{ptmri8t at 7pt}
\let\crnotice\mycrnotice%
\let\confname\myconfname%
\permission{Permission to make digital or hard copies of all or part of this work for personal or classroom use is granted without fee provided that copies are not made or distributed for profit or commercial advantage and that copies bear this notice and the full citation on the first page. Copyrights for components of this work owned by others than the author(s) must be honored. Abstracting with credit is permitted. To copy otherwise, or republish, to post on servers or to redistribute to lists, requires prior specific permission and/or a fee. Request permissions from Permissions@acm.org.}
\conferenceinfo{WSDM 2016}{February 22 - 25, 2016, San Francisco, CA, USA\\
{\mycrnotice{Copyright is held by the owner/author(s). Publication rights 
licensed to ACM.}}}
\copyrightetc{\the\acmcopyr}
\crdata{ACM 978-1-4503-3716-8/16/02/\$15.00 \\
DOI: http://dx.doi.org/10.1145/2835776.2835823}

\title{Personalized PageRank Estimation and Search: A Bidirectional Approach}
\date{\today}

\numberofauthors{3}
\author{
\alignauthor
Peter Lofgren\\
        \affaddr{Department of CS}\\
        \affaddr{Stanford University}\\
        \email{plofgren@cs.stanford.edu}
\alignauthor
Siddhartha Banerjee\\
        \affaddr{School of ORIE}\\
        \affaddr{Cornell University}\\
        \email{sbanerjee@cornell.edu}
\alignauthor
Ashish Goel\\
        \affaddr{Department of MS\&E}\\
        \affaddr{Stanford University}\\
        \email{ashishg@stanford.edu}
}

\usepackage[colorlinks,
            linkcolor=blue,
            citecolor=blue,
            urlcolor=magenta,
            linktocpage,
            plainpages=false,
            pdftex]{hyperref}

\begin{document}
\hypersetup{pageanchor=false}
\maketitle
\hypersetup{pageanchor=true}
\begin{abstract}
We present new algorithms for Personalized PageRank estimation and Personalized PageRank search.  
First, for the problem of estimating Personalized PageRank (PPR) from a source distribution to a target node, we present a new bidirectional estimator with simple yet strong guarantees on correctness and performance, and 3x to 8x speedup over existing estimators in experiments on a diverse set of networks.
Moreover, it has a clean algebraic structure which enables it to be used as a primitive for the Personalized PageRank Search problem: Given a network like Facebook, a query like ``people named \john,'' and a searching user, return the top nodes in the network ranked by PPR from the perspective of the searching user.
Previous solutions either score all nodes or score candidate nodes one at a time, which is prohibitively slow for large candidate sets.  
We develop a new algorithm based on our bidirectional PPR estimator which identifies the most relevant results by sampling candidates based on their PPR; this is the first solution to PPR search that can find the best results without iterating through the set of all candidate results.  
Finally, by combining PPR sampling with sequential PPR estimation and Monte Carlo, we develop practical algorithms for PPR search, and we show via experiments that our algorithms are efficient on networks with billions of edges.
\end{abstract}



\section{Introduction}
On social networks, personalization is necessary for returning relevant results for a query. For example, if a user searches for a common name like \john\ on a social network like Facebook, the results should depend on who is doing the search and who their friends are.  A good personalized model for measuring the importance of a node $t$ to a searcher $s$ is Personalized PageRank $\pi_s(t)$ \cite{Page1999,haveliwala2002topic,gupta2013wtf} -- this motivates a natural {\it Personalized PageRank Search Problem}: Given 
\begin{itemize}[nosep,leftmargin=*]
\item  a network with nodes $V$ (each associated with a set of keywords) and edges $E$ (possibly weighted and directed),
\item a keyword inducing a set of targets: \\
\centerline{$T = \{t \in V: \text{$t$ is relevant to the keyword}\}$}
\item a searching user $s \in V$ (or more generally, a distribution over starting nodes),
\end{itemize}
return the top-$k$ targets $t_1, \ldots, t_k \in T$ ranked by Personalized PageRank $\pi_s(t_i)$.

The importance of personalized search extends beyond social networks.
For example, personalized PageRank can be used to rank items in a
bi-partite user-item graph, in which there is an edge from a user to
an item if the user has liked that item.  This has proven useful on
YouTube when recommending videos \cite{baluja2008video} and on Twitter
for suggested users~\cite{Bahmani2010,gupta2013wtf}. On the web graph
there is a large body of work on using Personalized PageRank to rank
web pages (e.g.
\cite{Jeh2003,haveliwala2002topic}). 
The most clear-cut motivation for our work is for the social network name-search application discussed above, which we use as a running example in this paper.

The personalized search problem is difficult because every searching user has a different ranking on the target nodes. One naive solution would be to precompute the ranking for every searching user, but if our network has $n$ users this requires $\Theta(n^2)$ storage, which is clearly infeasible.  Another naive baseline would be to use power iteration~\cite{Page1999} at query time, but that would take $\Theta(m)$ computation between the search query and response, where $m$ is the number edges, which is also clearly infeasible. The challenge we face is to create a data structure much smaller than $O(n^2)$ which allows us to rank $\size{T}$ targets in response to a query in less than  $O(\size{T})$ time.

Previous work has considered the problem of personalized search on social networks.  For example Vieira et.~al.~\cite{vieira2007} consider this problem and provide excellent motivation for why results to a name-search query should be ranked based the friendships of the searching user and the candidate results.  They and others (e.g. \cite{Bahmani2012}) propose to rank results by shortest path length.  However, this metric doesn't take into account the number of paths between two users: If the searcher and two results \john\ A and \john\ B are distance 3 apart, but the searcher and \john\ A are connected by 100 length-3 paths while the searcher and \john\ B are connected by a single length-3 path, than \john\ A should be ranked above \john\ B, yet the shortest distance can't distinguish the two. 
To the best of our knowledge, no prior work has solved the Personalized PageRank search problem using less than $O(n^2)$ storage and $O(\size{T})$ query time.  The reason we are able to solve this is by exploiting a new bidirectional method of PageRank, introduced in \cite{fastppr} and improved in this work.

Our search algorithm is based on two key ideas.  The first is that we can find the top target nodes without having to consider each separately by \emph{sampling a target $t_i \in T$ in proportion to its Personalized PageRank $\pi_s(t_i)$}.  Because the top results typically have a much higher personalized PageRank than an average result, by sampling we can find the top results without iterating over all the results.  The second idea is that the probability of a random walk exactly reaching an element in $T$ is often very small, but by pre-computing an expanded set of nodes around each target, we can efficiently sample random walks until they get close to a target node, and then use the pre-computed data to sample targets $t_i$ in proportion to $\pi_s(t_i)$. 

There are currently two main limitations to our work.  First, because we do pre-computation on the set of nodes relevant to a query, we need the set of queries to be known in advance, although in the case of name search we can simply let the space of queries be the set of all first or last names.  Second, the pre-computed storage is significant; for name-search it is $O\pn{n  \sqrt{m}}$ to achieve query running time $O(\sqrt{m})$, where $n$ is the number of nodes and $m$ is the number of edges.  However, large graphs tend to be sparse, so this is still much smaller than $O\pn{n^2}$ and is less storage than any prior solution to the Personalized PageRank Search problem.  Also, pre-computation doesn't need to be done for all queries: for queries with small or very large target sets we describe alternative algorithms which do not require pre-computation.  These alternatives also overcome the limitation on queries being known in advance.



\noindent\textbf{Contributions:} To summarize, in this work we present:
\begin{itemize}[nosep,leftmargin=*]
\item A new bidirectional PageRank estimator, \texttt{Bidirectional-PPR} (section \ref{sec:bippr}), which has the following features:
  \begin{itemize}[nosep,leftmargin=*]
  \item \emph{Simple analysis}: We combine a simple linear-algebraic invariant with standard concentration bounds. The new analysis also allows generalizations to arbitrary Markov Chains, as done in \cite{generalized_fastppr}.
  \item \emph{Easy to implement}: The complete algorithm is only 18 lines of pseudo-code.  
  \item \emph{Significant empirical speedup}: For a given accuracy, it executes 3x-8x faster than the fastest previous algorithm, \texttt{FAST-PPR} \cite{fastppr}, on a diverse set of networks.
  \item \emph{Simple linear structure:} As shown in section \ref{sec:bippr_dot_product}, the estimates are a simple dot-product between a forward vector $x^s$ and a reverse vector $y^t$ -- this enables the development of PPR samplers.  
  \item \emph{Parallelizability:} Because the estimate is a dot-product, the precomputed vectors can be sharded across many servers, and the estimation algorithm can be naturally distributed, as shown in \cite{cross_partitioning}.
\end{itemize}
\item Two new solutions to the Personalized PageRank Search problem -- \texttt{BiPPR-Grouped} and \texttt{BiPPR-Sampling}. Given any set of targets $T$:
\begin{itemize}[nosep,leftmargin=*]
\item \bipprgrouped{} precomputes and stores the reverse vectors $y^t, t\in T$ after grouping them by their coordinates. This exploits the natural sparsity of these vectors to speed-up the computation of the PPR estimates at runtime.  
\item \bipprsampling{} samples nodes $t\in T$ proportional to their PPR $\pi_s(t)$. Now since PPR values are usually highly skewed, this serves as a good proxy for finding the top $k$ search results. 
\end{itemize}
\item Extensive simulations on the Twitter-2010 network to test the scalability of our algorithms for PPR-search. 
Our experiments demonstrate the trade-off between storage and runtime, and suggest that we should use a combination of methods, depending on the size of the set of targets $T$ induced by the keyword.
\end{itemize}


\section{Preliminaries}

We are given a graph $G=(V, E)$ with $n$ nodes and $m$ edges.  Define the out-neighbors of a node $u$ by $\outneighbors{u}=\{v: (u, v) \in E\}$ and let $\outdegree{u} = \size{\outneighbors{u}}$; define $\inneighbors{u}$ and $\indegree{u}$ similarly.  Define the average degree of nodes $\bar{d} = \frac{m}{n}$.  If the graph is weighted, for each $(u, v) \in E$ there is some positive weight $w_{u,v}$; otherwise we define $w_{u,v}=\frac{1}{\outdegree{u}}$ for all $(u, v) \in E$.  For simplicity we assume the weights are normalized such that for all $u$, $\sum_v w_{u,v} = 1$.

The personalized PageRank from source distribution $\sigma$ to target node $t$ can be defined using linear algebra as the solution to the equation
$ \pi_\sigma = \pi_\sigma (\alpha \sigma + (1 - \alpha) W) $, or equivalently defined using random walks
\begin{align*}
  \pi_\sigma(t) = \Pr[&\text{a random walk starting from $s \sim \sigma$} \\   
&\text{of length $\sim$ geometric($\alpha$) stops at $t$}]
\end{align*}
as shown in \cite{Avrachenkov2007}.  For concreteness, in this paper we often assume $\sigma = e_s$ for some single node $s$ (meaning the random walks always start at a single node $s$), but all results extend in a straightforward manner to any starting distribution $\sigma$.

Personalized PageRank was first defined in the original PageRank paper \cite{Page1999}.  For more on the motivation of Personalized PageRank, see \cite{haveliwala2002topic} and the survey \cite{Gleich-preprint-pagerank-beyond}.

\section{PageRank Estimation}
\label{sec:bippr}

In this section, we present our new bidirectional algorithm for PageRank estimation. We first develop the basic algorithm along with its theoretical performance guarantees; next, we outline some extensions of the basic algorithm; finally, we conclude the section with simulations demonstrating the efficiency of our technique. 

\noindent\textbf{The \texttt{Bidirectional-PPR} Algorithm} 

At a high level, our algorithm estimates $\pi_s(t)$ by first working backwards from $t$ to find a set of intermediate nodes `near' $t$ and then generating random walks forwards from $s$ to detect this set.

The reverse work from $t$ is done via the \rpush{} algorithm (see Algorithm \ref{alg:invPPR}) of Andersen et.~al.~\cite{Andersen2007}, that, given a target $t$ and a desired \emph{additive error-bound} $\epr$, produces estimates $p^t(s)$ of the PPR  $\pi_s(t)$ for every start node $s$.
More specifically, the \rpush{} algorithm produces two non-negative vectors $p^t \in \R^n$ and $r^t \in \R^n$ which satisfy the following invariant (Lemma $1$ in \cite{Andersen2007})
\begin{equation}
  \label{eq:ppr_dot_product}
  \pi_s(t) = p^t(s) + \sum_{v \in V} \pi_s(v) r^t(v).
\end{equation}
\rpush{} terminates once each residual value $r^t(v) < \epr$; now, viewing $\sum_{v \in V} \pi_s(v) r^t(v)$ as an error term, Andersen et al. observe that $p^t(s)$ estimates $\pi_s(t)$ up to a maximum additive error of $\epr$.

Our \texttt{Bidirectional-PPR} algorithm is based on the observation that in order to estimate $\pi_s(t)$ for a \emph{particular} $(s, t)$ pair, we can boost the accuracy by sampling and adding the residual values $r^t(v)$ from nodes $v$ which are sampled from $\pi_s$.
To see this, we first interpret Equation \eqref{eq:ppr_dot_product} as an expectation:
\[  \pi_s(t) = p^t(s) + \EE_{v \sim \pi_s}[r^t(v)]. \]
Now, since $\max_v r^t(v) < \epr$, the expectation $\EE_{v \sim \pi_s(v)}[r^t(v)]$ can be efficiently estimated using Monte Carlo.  
To do so, we generate $w = c \frac{\epr}{\delta}$ random walks of length $Geometric(\alpha)$ from start node $s$; here $c$ is a parameter which depends on the desired accuracy, $\epr$ is the maximum residual after running \rpush{}, and $\delta$ is the minimum PPR value we want to accurately estimate.  
Let $V_i$ be the final node of the $i^{th}$ random walk; note that $\Pr[V_i=v] = \pi_s(v)$.  
Let $X_i = r^t(V_i)$ denote the residual from the final node of the $i$th random walk, and $\bar{X} = \frac{1}{w} \sum_{i=1}^w X_i$. 
Then \texttt{Bidirectional-PPR} returns as an estimate of $\pi_s(t)$: 
\[\widehat{\pi}_s(t) = p^t(s) + \bar{X}
\]
The complete pseudocode is given in Algorithm \ref{alg:BIPPR}.

\begin{algorithm}[!ht]
\caption{\rpush{}$(G, \alpha, t,\epr)$~\cite{Andersen2007}}
\label{alg:invPPR}
\begin{algorithmic}[1]
\REQUIRE graph $G$ with edge weights $w_{u,v}$, teleport probability $\alpha$, target node $t$, maximum residual $\epr$
\STATE Initialize (sparse) estimate-vector $p_t = \vec{0}$ and (sparse) residual-vector $r_t = e_t$ (i.e.~$r_t(v) = 1$ if $v=t$; else $0$)

\WHILE{$\exists v\in V\,s.t.\,r_t(v)> \epr$}
\FOR{$u\in \inneighbors{v}$}
   \STATE $r_t(u) \pluseq (1 - \alpha) w_{u,v} r_t(v)$
\ENDFOR
\STATE      $p_t(v) \pluseq \alpha r_t(v)$
\STATE      $r_t(v)=0$
\ENDWHILE
\RETURN $(p_t, r_t)$
\end{algorithmic}
\end{algorithm}    

\begin{algorithm}[!ht]
\caption{\texttt{Bidirectional-PPR}$(s,t,\delta)$}
\label{alg:BIPPR}
\begin{algorithmic}[1] 
\REQUIRE graph $G$, teleport probability $\alpha$, start node $s$, target node $t$, minimum probability $\delta$, accuracy parameter $c$ (in our experiments we use $c = 7$)
\STATE Choose $\epr = c_{\text{balance}}/\sqrt{m})$, where $c_{\text{balance}}$ is tuned to balance forward and reverse work.  (For greater efficiency, use the balanced version described in Section \ref{sec:balance}.)
\STATE $(p_t, r_t)$ =  \texttt{Approx-Contributions}$(t,\epr, \alpha)$

\STATE Set number of walks $w=c \epr/\delta$ \quad(cf. Theorem \ref{thm:bidirmain})
\FOR{index $i\in [w]$}
\STATE Sample a random walk starting from $s$ (sampling a start from $s$ if $s$ is a distribution), stopping after each step with probability $\alpha$; let $v_i$ be the endpoint
\STATE Set $X_i = r_t(v_i)$
\ENDFOR 
\RETURN $\widehat{\PR}_s(t)=p_t(s) + (1/w)\sum_{i\in[w]}X_i$
\end{algorithmic}
\end{algorithm}

\noindent\textbf{Accuracy Analysis} 


We first prove that \texttt{Bidirectional-PPR} returns an estimate with the desired accuracy with high probability:  

\begin{theorem}
\label{thm:bidirmain} 
Given start node $s$ (or source distribution $\sigma$), target $t$, minimum PPR $\delta$, maximum residual $\epr > \frac{2e \delta}{\alpha \epsilon}$, relative error $\epsilon \leq 1$, and failure probability $\pfail$, \bippr{} outputs an estimate $\widehat{\PR}_s(t)$ such
that with probability at least $1- \pfail$ the following hold: 
\begin{itemize}[nosep,leftmargin=*]
\item If $\pi_s(t) \geq \delta$: \hspace{1cm} $\abs{\pi_s(t)-\hat{\pi}_s(t)} \leq \epsilon \pi_s(t)$. 
\item If $\pi_s(t) \leq \delta$: \hspace{1cm}
$\abs{\pi_s(t)-\hat{\pi}_s(t)} \leq 2e\delta$.
\end{itemize}
\end{theorem}
The above result shows that the estimate $\hat{\pi}_s(t)$ can be used to distinguish between `significant' and `insignificant' PPR pairs: for pair $(s,t)$, Theorem \ref{thm:bidirmain} guarantees that if $\pi_s(t) \geq \frac{(1+2e)\delta}{(1-\epsilon)}$, then the estimate is greater than $(1+2e)\delta$, whereas if $\pi_s(t) < \delta$, then the estimate is less than $(1+2e)\delta$. 
The assumption $\epr > \frac{2e \delta}{\alpha \epsilon}$ is easily satisfied, as typically $\delta=O\pn{\frac{1}{n}}$ and $\rmax =\Omega\pn{\frac{1}{\sqrt{m}}}$.  

\begin{proof} 
As shown in Algorithm \ref{alg:BIPPR}, we will average over 
\[ w = c \frac{\epr}{\delta} \]
walks, where $c$ is a parameter we choose later.
Each walk is of length $Geometric(\alpha)$, and we denote $V_i$ as the last node visited by the $i^{th}$ walk, so that $V_i\sim\pi_s$.
Let $X_i = r^t(V_i)$.  The estimate returned by \texttt{Bidirectional-PPR} is
\[\widehat{\pi}_s(t) = p^t(s) + \frac{1}{w} \sum_{i=1}^w X_i. \]
First, from Equation \eqref{eq:ppr_dot_product}, we have that $\EE[\widehat{\pi}_s(t)]= \pi_s(t)$.   
Moreover, \rpush{} guarantees that for all $v$, $r^t(v)<\epr$, and so each $X_i$ is bounded in $[0,\epr]$.  Before applying Chernoff bounds, we rescale $X_i$ by defining $Y_i = \frac{1}{\epr} X_i \in [0, 1]$, and we define $Y = \sum_{i=1}^w Y_i$.

We will show concentration of the estimates via the following two Chernoff bounds (see Theorem $1.1$ in~\cite{DuPa09}):
\begin{enumerate}
\item $\PP[|Y - \EE[Y]| > \epsilon \EE[Y]] < 2 \exp(-\frac{\epsilon^2}{3}\EE[Y])$
\item $\textrm{For any } b > 2e\EE[Y], \PP[Y > b] \leq 2^{-b}$
\end{enumerate}
We perform a case analysis based on whether $\EE[X_i] \geq \delta$ or $\EE[X_i] < \delta$. 

First suppose $\EE[X_i] \geq \delta$.  This implies that $\pi_s(t) \geq \delta$ so we will prove a relative error bound of $\epsilon$.
Now we have $\EE[Y] = \frac{w}{\epr} \EE[X_i] = \frac{c}{\delta} \EE[X_i] \geq c$, and thus:
\begin{align*}
\PP[\abs{\widehat{\pi}_s(t) - \pi_s(t)} > \epsilon \pi_s(t)] 
 &\leq \PP[\abs{\bar{X} - \EE[X_i]} > \epsilon \EE[X_i]] \\
  &= \PP[\abs{Y - \EE[Y]} > \epsilon \EE[Y]] \\
  &\leq 2 \exp\pn{-\frac{\epsilon^2}{3}\EE[Y]} \\
  &\leq 2 \exp\pn{-\frac{\epsilon^2}{3} c}
  \leq \pfail,
\end{align*}
where the last line holds as long as we choose
\[ c \geq \frac{3}{\epsilon^2} \ln \pn{\frac{2}{\pfail}}.\]

Suppose alternatively that $\EE[X_i] < \delta$.  Then
\begin{align*}
\PP[\abs{\hat{\pi}_s(t) - \pi_s(t)} > 2e\delta]
&= \PP[\abs{\bar{X} - \EE[X_i]} > 2e\delta] \\
&=  \PP\bk{\abs{Y -  \EE[Y]} > \frac{w}{\epr} 2e\delta} \\
&\leq  \PP\bk{Y > \frac{w}{\epr}2e\delta} .
\end{align*}
At this point we set $b = \frac{w}{\epr} 2e\delta=2ec$ and apply the second Chernoff bound.  Note that $\EE[Y] = \frac{c}{\delta} \EE[X_i] < c$, and hence we satisfy $b > 2e \EE[Y]$.
The second bound implies that
\begin{equation}
  \label{eq:1}
   \PP[\abs{\hat{\pi}_s(t) - \pi_s(t)} > 2e\delta] \leq 2^{-b} \leq \pfail
\end{equation}
as long as we choose $c$ such that:
\[ c \geq \frac{1}{2e} \log_2 \frac{1}{\pfail}. \]
If  $\pi_s(t) \leq \delta$, then equation \ref{eq:1} completes our proof. 

The only remaining case is when $\pi_s(t) > \delta$ but $\EE[X_i] < \delta$. This implies that $p^t(s) > 0$ since $\pi_s(t) = p^t(s) + \EE[X_i]$.  In the \rpush{} algorithm when we increase $\pi_s(t)$, we always increase it by at least $\alpha \rmax$, so we have $p^t(s) \geq \alpha \rmax$.  We have that
\[ \frac{\abs{\hat{\pi}_s(t) - \pi_s(t)}}{\pi_s(t)} \leq \frac{\abs{\hat{\pi}_s(t) - \pi_s(t)}}{\alpha \rmax}. \]
By assumption, $\frac{2e \delta}{\alpha \rmax} < \epsilon$, so by equation \ref{eq:1},
\[ \PP\bk{\frac{\abs{\hat{\pi}_s(t) - \pi_s(t)}}{\pi_s(t)} > \epsilon} \leq \pfail \]

The proof is completed by combining all cases and choosing $c = \frac{3}{\epsilon^2} \ln \pn{\frac{2}{\pfail}}$.
We note that the constants are not optimized; in experiments we find that $c=7$ gives mean relative error less than 8\% on a variety of graphs.
\end{proof}

\noindent\textbf{Running Time Analysis}

The runtime of \texttt{Bidirectional-PPR} depends on the target $t$: if $t$ has many in-neighbors and/or large global PageRank $\pi(t)$, then the running time will be slower than for a random $t$. Theorem 1 of \cite{Andersen2007} states that \rpush{} ($G, \alpha, t, \epr$) performs $\frac{ n \pi(t)}{\alpha \epr}$ pushback operations, and the exact running time is proportional to the sum of the in-degrees of all the nodes where we pushback from.  In the worst case, we might have $\indegree{t} = \Theta(n)$ and \texttt{Bidirectional-PPR} takes $\Theta(n)$ time.  
However, for \emph{a uniformly chosen target node}, we can prove the following:
\begin{theorem}
\label{thm:fastpprtime} 
For any start node $s$ (or source distribution $\sigma$), minimum PPR $\delta$, maximum residual $\epr$, relative error $\epsilon$, and failure probability $\pfail$, if the target $t$ is chosen uniformly at random, then \emph{\texttt{Bidirectional-PPR}} has expected running time 
\[O\pn{ \sqrt{\frac{\bar{d}}{\delta}} \frac{\sqrt{\log\pn{1/\pfail}}}{\alpha \epsilon}}.
 \]
\end{theorem}

In contrast, the running time for \texttt{Monte-Carlo} to achieve the same accuracy guarantee is $O\pn{\frac{1}{\delta} \frac{\log\pn{1/\pfail}}{\alpha \epsilon^2}}$, and the running time for \rpush{} is $O\pn{\frac{\bar{d}}{\delta \alpha}}$.  The fastest previous algorithm for this problem, the \texttt{FAST-PPR} algorithm of \cite{fastppr}, has an average running time bound of
$O\pn{\frac{1}{\alpha\epsilon^2}\sqrt{\frac{\bar{d}}{\delta}} \sqrt{ \frac{\log\pn{1/\pfail} \log\pn{1/\delta}}{\log\pn{1/(1-\alpha)}} }}$
 for uniformly chosen targets.  
The running time bound of \texttt{Bidirectional-PPR} is thus asymptotically better than \texttt{FAST-PPR}, and in experiments the constants required for the same accuracy are smaller, making \texttt{Bidirectional-PPR} is 3 to 8 times faster on a diverse set of graphs.

\begin{proof} 
In \cite{Lofgren2013}, it is proven that for a uniform random $t$, Approx-Contributions runs in average time $\frac{\bar{d}}{\alpha \epr}$ where $\bar{d}$ is the average degree of a node. On the other hand, from Theorem  \ref{thm:bidirmain}, we know that we need to generate
$O\pn{ \frac{\epr}{\delta\epsilon^2} \ln\pn{1/\pfail}}$ random walks, each of which can be sampled in average time $1/\alpha$.  Finally, we choose $\epr = \frac{\epsilon}{\alpha} \sqrt{\frac{\bar{d}}{ \ln\pn{2/\pfail}}}$ to minimize our running time bound and get the claimed result. 
\end{proof}

\noindent\textbf{Extensions}
\label{sec:balance}
Bidirectional-PageRank extends naturally to generalized PageRank using a source distribution $\sigma$ rather than a single start node -- we simply sample an independent starting node for each walk, and replace $p_t(s)$ with the expected value of $p_t(s)$ when $s$ is sampled from the starting distribution.

The dynamic runtime-balancing method proposed in \cite{fastppr} can improve the running time of Bidirectional-PageRank in practice.  In this technique, $\epr$ is chosen dynamically in order to balance the amount of time spent by \rpush{} and the amount of time spent generating random walks. 
To implement this, we modify \rpush{}  to use a priority queue in order to always push from the node $v$ with the largest value of $r_t(v)$.  
We also change the while loop so that it terminates when the amount of time spent achieving the current value of $\epr$ first exceeds the predicted amount of time required for sampling random walks, $c_{\text{walk}}  \cdot c \cdot \frac{\epr}{\delta}$, where $c_{\text{walk}}$ is the average time it takes to sample a random walk.  For full pseudocode, see \cite{peterThesis}.

\noindent\textbf{Experimental Validation}

\begin{figure*}[t]
\centering
\subfigure[Sampling targets uniformly]{
\label{fig:runtime_uniform}
\includegraphics[width=0.9\columnwidth]{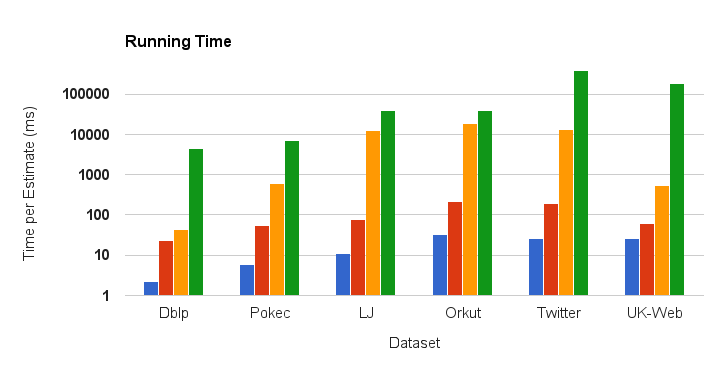}
}
\hfill
\subfigure[Sampling targets from PageRank distribution]{
\label{fig:runtime_pagerank}
\includegraphics[width=1.1\columnwidth]{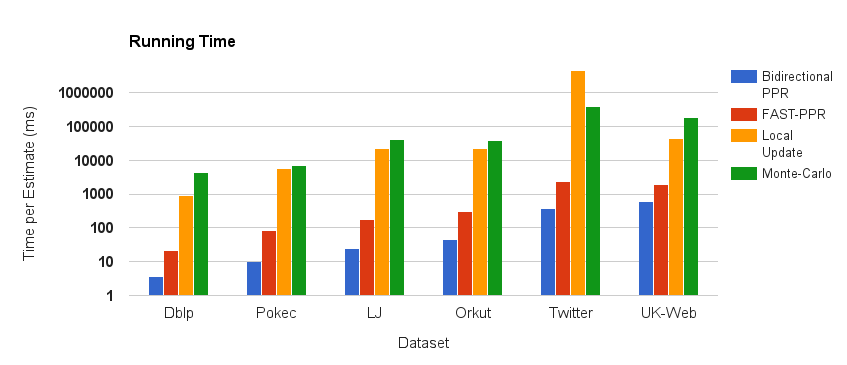}
}
\caption[Running-time Plots]{Average running-time (on log-scale) for different networks. We measure the time required for estimating PPR values $\pi_s(t)$ with threshold $\delta=\frac{4}{n}$ for $1000$ $(s,t)$ pairs. For each pair, the start node is sampled uniformly, while the target node is sampled uniformly in Figure \ref{fig:runtime_uniform}, or from the global PageRank distribution in Figure \ref{fig:runtime_pagerank}. In this plot we use teleport probability $\alpha=0.2$.}
\label{fig:ppr_runtime}
\end{figure*}

We now compare \texttt{Bidirectional-PPR} to its predecessor algorithms (namely: \texttt{FAST-PPR} \cite{Lofgren2013}, \texttt{Monte Carlo} \cite{Avrachenkov2007,Fogaras2005} and \rpush{} \cite{Andersen2007}).
The experimental setup is identical to that in \cite{Lofgren2013}; for convenience, we describe it here in brief. 
We perform experiments on 6 diverse, real-world networks: two directed social networks (Pokec (31M edges) and Twitter-2010 (1.5 billion edges)), two undirected social network (Live-Journal (69M edges) and Orkut (117M edges)), a collaboration network (dblp (6.7M edges)), and a web-graph (UK-2007-05 (3.7 billion edges)).  
Since all algorithms have parameters that enable a trade-off between running time and accuracy, we first choose parameters such that the mean relative error of each algorithm is approximately 10\%.  
For bidirectional-PPR, we find that setting $c=7$ (i.e., generating $7 \cdot \frac{\epr}{\delta}$ random walks) results in a mean relative error less than 8\% on all graphs; for the other algorithms, we use the settings determined in \cite{Lofgren2013}. 
We then repeatedly sample a uniformly-random start node $s \in V$, and a random target $t \in T$ sampled either uniformly or from PageRank (to emphasize more important targets).  
For both \texttt{Bidirectional-PPR} and \texttt{FAST-PPR}, we used the dynamic-balancing heuristic described above.
The results are shown in Figure \ref{fig:ppr_runtime}.  

Note that \texttt{Bidirectional-PPR} is $3$ to $8$ times faster than \texttt{FAST-PPR} across all graphs.
In particuar, \texttt{Bidirectional-PPR} only needs to sample $7 \frac{\epr}{\delta}$ random walks, while FAST-PPR needs $350 \frac{\epr}{\delta}$ walks to achieve the same mean relative error.  
This is because \texttt{Bidirectional-PPR} is unbiased, while \texttt{FAST-PPR} has a bias from \rpush{}.



\section{Personalized PageRank Search}

We now turn from Personalized PageRank estimation to the Personalized PageRank search problem: 
\begin{center}
\emph{Given a start node $s$ (or distribution $\sigma$) and a query $q$ which filters the set of all targets to some list $T = \{t_i\} \subseteq V$, return the  top-$k$ targets ranked by $\pi_s[t_i]$.}
\end{center}

We consider as baselines two algorithms which require no pre-computation.  They are efficient for certain ranges of $\size{T}$, but our experiments show they are too slow for real-time search across most values of $\size{T}$:
\begin{itemize}[nosep,leftmargin=*]
\item \noindent\textbf{\mc{}}~\cite{Avrachenkov2007,Fogaras2005}:  Sample random walks from $s$, and filter out any walk whose endpoint is not in $T$.  If we desire $\nsample$ samples, this takes time $O\pn{\nsample/\pi_s[T]}$, where $\pi_s[T] := \sum_{t \in T} \pi_s[t]$ is the probability that a random walk terminates in $T$.  
This method works well if $T$ is large, but in our experiments on Twitter-2010 it takes minutes per query for $\size{T}=1000$ (and hours per query for $\size{T}=10$).

\item \noindent\textbf{\bipprbasic{}}: On the other hand, we can estimate $\pi_s[t]$ to each $t \in T$ separately using \bippr.  
This has an average-case running time $O\pn{\size{T} \sqrt{\bar{d}/\delta_k}}$ where $\delta_k$ is the PPR of the $k^{th}$ best target.  
This method works well if $T$ is small, but is too slow for large $T$; 
in our experiments, it takes on the order of seconds for $\size{T} \leq 100$, but more than a minute for $\size{T}=1000$.
\end{itemize}

If we are allowed pre-computation, then we can improve upon \bippr{} 
by precomputing and storing a reverse vector from all target nodes. 
\label{ppr_as_dot_product} To this end, we first observe that the estimate $\hat{\pi}_s[t]$  can be written as a dot-product.
Let $\tilde{\pi}_s$ be the empirical distribution over terminal nodes due to $w$ random walks from $s$ (with $w$ chosen as in Theorem \ref{thm:bidirmain}); we define the \emph{forward vector} $x_s \in \R^{2n}$ to be the concatenation of the basis vector $e_s$ and the random-walk terminal node distribution.
On the other hand, we define the \emph{reverse vector} $y^t \in \R^{2n}$, to be the concatenation of the estimates $p^t$ and the residuals $r^t$.
Formally, define
\begin{equation}
\label{eq:frvectors}
x_s = (e_s, \tilde{\pi}_s) \in \R^{2n},\quad\quad y^t = (p^t, r^t) \in \R^{2n}.
\end{equation}  
Now, from Algorithm \ref{alg:BIPPR}, we have
\begin{equation}
\label{eq:dotprod}
	\hat{\pi}_s[t] = \langle x_s,y^t\rangle .
\end{equation}

The above observation motivates the following algorithm:
\begin{itemize}[nosep,leftmargin=*]
\item \noindent\textbf{\bipprprecomp{}}: In this approach, we first use \rpush{} to pre-compute and store a reverse vector $y^t$ for each $t \in V$. At query time, we generate random walks to form the forward vector $x_s$; now, given any set of targets $T$, we compute $\size{T}$ dot-products $\langle x_s, y^t\rangle$, and use these to rank the targets.  
This method now has an \emph{worst-case} running time $O\pn{\size{T} \sqrt{\bar{d}/\delta_k}}$. 
In practice, it works well if $T$ is small, but is too slow for large $T$. In our experiments (doing 100,000 random walks at runtime) this approach takes around a second for $\size{T} \leq 30$, but this climbs to a minute for $\size{T}=10,000$.
\end{itemize} 

The \bipprprecomp{} approach is faster than \bipprbasic{} (at the cost of additional precomputation and storage), and also faster than \mc{} for small sets $T$, but it is still not efficient enough for real-time personalized search.  
This motivates us to find a more efficient algorithm that scales better than \bipprbasic{} for large $T$, yet is fast for small $|T|$. 
In the following sections, we propose two different approaches for this -- the first based on pre-grouping the precomputed reverse-vectors, and the second based on sampling target nodes from $T$ according to PPR.  
For convenience, we first summarize the two approaches: 

\begin{itemize}[nosep,leftmargin=*]
\item \noindent\textbf{\bipprgrouped{}}: Here, as in \bipprprecomp{}, we compute an estimate to each $t \in T$ using \bippr{}. 
However, we leverage the sparsity of the reverse vectors $y^t = (p^t, r^t)$ by first grouping them in a way we will describe. This makes the dot-product more efficient.   
This method has a \emph{worst-case} running time of $O\pn{\size{T} \sqrt{\bar{d}/\delta_k}}$, and in experiments we find it is much faster than \bipprprecomp{}.  
For our parameter choices its running time is less than 250ms across the range of $\size{T}$ we tried.

\item \noindent\textbf{\bipprsampling{}}: We again first pre-compute the reverse vectors $y^t$. 
Next, for a given target $t$, we define the \emph{expanded target-set} $T_t=\{v \in [2n]|y^t[v]\neq 0\}$, i.e., the set of nodes with non-zero reverse vectors from $t$. 
At run-time, we now sample random walks forward from $s$ to nodes in the expanded target sets. 
Using these, we create a sampler in average time $O\pn{\epr/\delta_k}$ (where as before $\delta_k$ is the $k^{th}$ largest PPR value $\pi_s[t_k]$), which samples nodes $t \in T$ with probability proportional to the PPR $\pi_s[t]$.  We describe this in detail in Section \ref{sec:sampler}. 
Once the sampler has been created, it can be sampled in $O(1)$ time per sample.  
The algorithm works well for any size of $T$, and has the unique property that in can identify the top-$k$ target nodes without computing a score for all $\size{T}$ of them.  For our parameter choice its running time is less than 250ms across the range of $\size{T}$ we tried.
\end{itemize}

We note here that the case $k=1$ (i.e., for finding the top PPR node) corresponds to solving a Maximum Inner Product Problem.  
In a recent line of work, Shrivastava and Li~ \cite{shrivastava2014asymmetric,shrivastava2014improved} propose a sublinear time algorithm for this problem based on Locality Sensitive Hashing; however, their method assumes that there is some bound $U$ on $\norm{y^t}_2$ and that $\max_t \langle x_s , y^t\rangle$ is a large fraction of $U$. 
In personalized search, we usually encounter small values of $\max_t \langle x_s , y^t\rangle$ relative to $\max \norm{y^t}_2$ -- finding an LSH for Maximum Inner Product Search in this regime is an interesting open problem for future research.  
Our two approaches bypass this by exploiting particular structural features of the problem -- \bipprgrouped{} exploits the sparsity of the reverse vectors to speed up the dot-product, and \bipprsampling{} exploits the skewed distribution of PPR scores to find the top targets without even computing full dot-products.


\subsection{Bidirectional-PPR with Grouping} 
\label{sec:bippr_dot_product} 

In this method we improve the running-time of \bipprprecomp{} by pre-grouping the reverse vectors corresponding to each target set $T$.
Recall that in \bipprprecomp{}, we first pre-compute reverse vectors $y^t = (p^t, r^t) \in \R^{2n}$ using \rpush{} for each $t$.  At run-time, given $s$, we compute forward vector $x_s = (e_s, \tilde{\pi}_s)$ by generating sufficient random-walks, and then compute the scores $\langle x_s , y^t\rangle$ for $t \in T$.  
Our main observation is that \emph{we can decrease the running time of the dot-products by pre-grouping the vectors $y^t$ by coordinate}.  
The intuition behind this is that in each dot product $\sum_v x_s[v] y^t[v]$, the nodes $v$ where $x_s[v] \neq 0$ often don't have $y^t[v] \neq 0$, and most of the product terms are 0.   
Hence, we can improve the running time by grouping the vectors $y^t$ in advance by coordinate $v$.  Now, at run-time, for each  $v$ such that $x_s[v] \neq 0$, we can efficiently iterate over the set of targets $t$ such that $y^t[v] \neq 0$.  

An alternative way to think about this is as a sparse matrix-vector multiplication $Y^{T} x_s$ after we form a matrix $Y^T$ whose rows are $y^t$ for $t \in T$.  
This optimization can then be seen as a sparse column representation of that matrix.

\begin{algorithm}[ht]
\caption{BiPPRGroupedPrecomputation$(T, \epr)$}
\label{alg:approxContributionsToSet}
\begin{algorithmic}[1]
\REQUIRE Graph $G$, teleport probability $\alpha$, target nodes $T$, maximum residual $\epr$
\STATE $z\leftarrow$ empty hash map of vectors such that for any $v$, $z[v]$ defaults to an empty (sparse) vector in $\R^{2|V|}$
\FOR{$t \in T$}
  \STATE Compute $y^t = (p_t, r_t)\in \R^{2|V|}$ via \rpush{}$(G, \alpha, t, \epr)$
  \FOR{$v \in [2 \size{V}] \text{ such that } y_t[v] > 0$}
    \STATE $z[v][t] = y^t[v]$
  \ENDFOR
\ENDFOR
\RETURN $z$
\end{algorithmic}
\end{algorithm}
\vspace{-0.2cm}

\begin{algorithm}[ht]
\caption{BiPPRGroupedRankTargets$(s, \epr, z)$}
\label{alg:RankTargets}
\begin{algorithmic}[1] 
\REQUIRE Graph $G$, teleport probability $\alpha$, start node $s$, maximum residual $\epr$, $z$: hash map of reverse vectors grouped by coordinate 
\STATE Set number of walks $w = c \frac{\epr}{\delta}$  (In experiments we found $c=20$ achieved precision@3 above 90\%.)
\STATE Sample $w$ random-walks of length $Geometric(\alpha)$ from $s$; compute $\tilde{\pi}_s[v]=$ fraction of walks ending at node $v$
\STATE Compute $x_s = (e_s, \tilde{\pi}_s) \in \R^{2|V|}$
\STATE Initialize empty map score from $V$ to $\R$
\FOR{$v$ such that $x_s[v] > 0$}
  \FOR{$t$ such that $z[v][t] > 0$}
    \STATE  score($t$) $\pluseq x_s[v] z_v[t]$
  \ENDFOR
\ENDFOR
\STATE Return $T$ sorted in decreasing order of score
\end{algorithmic}
\end{algorithm}

We refer to this method as \bipprgrouped{}; the complete pseudo-code is given in Algorithm \ref{alg:RankTargets}.  
The correctness of this method follows again from Theorem \ref{thm:bidirmain}.  
In experiments, this method is efficient for $T$ across all the sizes of $T$ we tried, taking less than 250 ms even for $\size{T} = 10,000$.  
The improved running time of \bipprgrouped{} comes at the cost of more storage compared to \bipprprecomp{}. 
In the case of name search, where each target typically only has a first and last name, each vector $y^t$ only appears in two of these pre-grouped structures, so the storage is only twice the storage of \bipprprecomp{}.
On the other hand if a target $t$ contains many keywords, $y^t$ will be included in many of these pre-grouped data structures, and storage cost will be significantly greater than for \bipprprecomp{}.



\subsection{Sampling from Targets Matching a Query}
\label{sec:sampler}

The key idea behind this alternate method for PPR-search is that by sampling a target $t$ in proportion to its PageRank we can quickly find the top targets without iterating over all of them.  
After drawing many samples, the targets can be ranked  according to the number of times they are sampled.  
Alternatively a full \bippr{} query can be issued for some subset of the targets before ranking. 
This approach exploits the skewed distribution of PPR scores in order to find the top targets.  
In particular, prior empirical work has shown that on the Twitter graph, for each fixed $s$, the values $\pi_s[t]$ follow a power law \cite{Bahmani2010}.

We define the PPR-Search Sampling Problem as follows:
\begin{center}
\emph{Given a source distribution $s$ and a query $q$ which filters the set of all targets to some list $T = \{t_i\} \subseteq V$, sample a target $t_i$ with probability $p[t_i] = \frac{\pi_s[t_i]}{\sum_{t \in T} \pi_s[t]}$.}
\end{center}
We develop two solutions to this sampling problem.  The first, in $O(w) = O(\epr/\delta_k)$ time, generates a data structure which can generate an arbitrary number of independent samples from a distribution which approximates the correct distribution. 
The second can generate samples from the exact distribution $\pi_s[t_i]$, and generates complete paths from $s$ to some $t \in T$, but requires time $O(\epr/\pi_s[T])$ per sample.  Because the approximate sampler is more efficient, we present that here and defer the exact sampler to \cite{peterThesis}. 

\noindent\textbf{The \bipprsampling{} Algorithm}

\begin{figure}[thb]
\centering
\includegraphics[width=1\columnwidth]{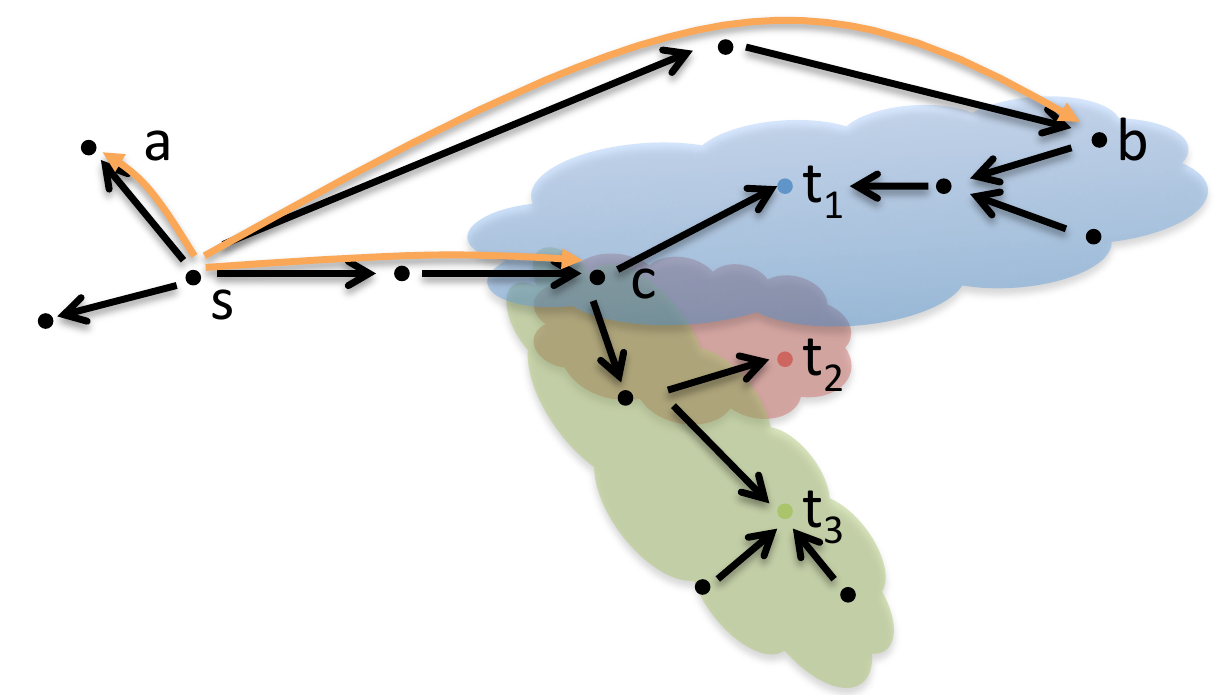}
\caption[Example]{Search Example: Given target set $T=\{t_1, t_2, t_3\}$, for each target $t_i$ we have drawn the expanded target-set, i.e., nodes $v$ with positive residual $y^{t_i}[v]$.  
From source $s$, we sample three random walks, ending at nodes $a$, $b$, and $c$.  
Now suppose $y^{t_1}(b)=0.64, y^{t_1}(c) = 0.4, y^{t_2}(c)=0.16$, and $y^{t_3}(c)=0.16$ -- note that the remaining residuals are $0$.
Then we have $y^T(a)=0, y^T(b)=0.64$ and $y^T(c)=0.72$, and consequently, the sampling weights of $(a,b,c)$ are $(0,0.213,0.24)$.    
Now, to sample a target, we first sample from $\{a,b,c\}$ in proportion to its weight. Then if we sample $b$, we always return $t_1$; if we sample $c$, we sample $(t_1,t_2,t_3)$ with probability $(5/9,2/9,2/9)$.}
\label{fig:example1}
\end{figure}

The high level idea behind our method is \emph{hierarchical sampling}.  
Recall that the start node $s$ has an associated forward vector $x_s = (e_s, \pi_s)$, and from each target $t$ we have a reverse vector $y^t$; the PPR-estimate is given by $\pi_s[t] \approx \langle x_s , y^t\rangle$.  
Thus we want to sample $t \in T$ with probability:
\[ p[t] = \frac{\langle x_s , y^t\rangle}{\sum_{j \in T} \langle x_s , y^{j}\rangle}. \]
We will sample $t$ in two stages: first we sample an intermediate node $v \in V$ with probability:
\[ p'_s[v] = \frac{x_s[v]  \sum_{j \in T}  y^{j}[v]}{\sum_{j \in T} \langle x_s ,  y^{j}\rangle}. \]
Following this, we sample $t\in T$ with probability:
\[ p''_v[t] = \frac{y^t[v]}{\sum_{j \in T} y^{j}[v]} . \]
It is easy to verify that $p[t]=\sum_{v\in V}p'_s[v]p''_v[t]$. 
Figure \ref{fig:example1} shows how the sampling algorithm works on an example graph.  The pseudo-code is given in Algorithm \ref{alg:samplerPrecompute} and Algorithm \ref{alg:SampleApprox}.

Note that we assume that some set of supported queries is known in advance, and we first pre-compute and store a separate data-structure for each query $Q$ (i.e., for each target-set $T= \{t \in V: t \text{ is relevant to } Q\}$).  In addition, we can optionally pre-compute $w$ random walks from each start-node $s$, and store the forward vector $x_s$, or we can compute $x_s$ at query time by sampling random walks.

%
\begin{algorithm}[ht]
\caption{SamplerPrecomputationForSet$(T, \epr)$}
\label{alg:samplerPrecompute}
\begin{algorithmic}[1]
\REQUIRE Graph $G$, teleport probability $\alpha$, target-set $T$, maximum residual $\epr$
\FOR{$t \in T$}
  \STATE Compute $y^t = (p^t, r^t)\in \R^{2\size{V}}$ via \rpush{}$(G, \alpha, t, \epr)$
\ENDFOR
\STATE Compute $y^{T} = \sum_{t \in T} y^t$
\FOR{$v \in V \text{ such that } y^T[v] > 0$}
  \STATE Create sampler$_v$ which samples $t$ with probability $p''_v[t]$  (For example, using the alias sampling method  \cite{Walker:1977:EMG:355744.355749}, \cite[section~3.4.1]{knuth1998vol2}).
\ENDFOR
\RETURN ($y^T, \{\text{sampler}_v\}$)
\end{algorithmic}
\end{algorithm}


\begin{algorithm}[ht]
\caption{SampleAndRankTargets$(s, \epr, y^T, \{\text{sampler}_v\})$}
\label{alg:SampleApprox}
\begin{algorithmic}[1] 
\REQUIRE Graph $G$, teleport probability $\alpha$, start node $s$, maximum residual $\epr$, reverse vectors $y^T$, intermediate-node-to-target samplers $\{\text{sampler}_v\}$.
\STATE Set number of walks $w = c \frac{\epr}{\delta}$. In experiments we found $c=20$ achieved precision@5 above 90\% on Twitter-2010.
\STATE Set number of samples $n_{s}$ (We use $n_{s} = w$)
\STATE Sample $w$ random walks from $s$ and let $\tilde{\pi}_s$ be the empirical distribution of their endpoints; compute forward vector $x_s = (e_s, \tilde{\pi}_s) \in \R^{2\size{V}}$
\STATE Create $\text{sampler}_s$ to sample $v \in [2 \size{V}]$ with probability $p'_s[v]$, i.e., proportional to $x_s[v] y^T[v]$
\STATE Initialize empty map score from $V$ to $\mathbb{N}$
\FOR{$j \in [0,n_{\text{s}}-1]$}
  \STATE Sample $v$ from $\text{sampler}_s$
  \STATE Sample $t$ from $\text{sampler}_v$
  \STATE Increment score($t$)
\ENDFOR
\STATE Return $T$ sorted in decreasing order of score
\end{algorithmic}
\end{algorithm}

\noindent\textbf{Running Time}: 
For a small relative error for targets  with $\pi_s[t] > \delta$, we use $w = c\epr/\delta$ walks, where $c$ is chosen as in Theorem \ref{thm:bidirmain}.  
The support of our forward sampler is at most $w$ so its construction time is $O(w)$ using the alias method of sampling from a discrete distribution \cite{Walker:1977:EMG:355744.355749},  \cite[section~3.4.1]{knuth1998vol2}. 
Once constructed, we can get independent samples in $O(1)$ time.  
Thus the query time to generate $\nsample$ samples is $O \pn{c \epr/\delta + \nsample}$. 

\noindent\textbf{Accuracy}:
\bipprsampling{} does not sample exactly in proportion to $\pi_s$; instead, the sample probabilities are proportional to a distribution $\hat{\pi}_s$ satisfying the guarantee of Theorem \ref{thm:bidirmain}. 
In particular, for all targets $t$ with $\pi_s[t] \geq \delta$, this will have a small relative error $\epsilon$, while targets with $\pi_s[t] < \delta$ will likely be sampled rarely enough that they won't appear in the set of top-$k$ most sampled nodes.

\noindent\textbf{Storage Required}: 
The storage requirements for \bipprsampling{}  (and for \bipprgrouped{}) depends on the distribution of keywords and how $\rmax$ is chosen for each target set.  For simplicity, here we assume a single maximum residual $\epr$ across all target sets, and assume each target is relevant to at most $\gamma$ keywords.  For example, in the case of name search, each user typically has a first and last name, so $\gamma = 2$.
\begin{theorem}
  \label{thm:storage_all_targets}
Let graph $G$, minimum-PPR value $\delta$ and time-space trade-off parameter $\rmax$ be given, and suppose every node contains at most $\gamma$ keywords. Then the total storage needed for \bipprsampling{} to construct a sampler for any source node (or distribution) $s$ and any set of targets $T$ corresponding to a single keyword is $O\pn{\frac{\gamma m}{\alpha \epr}}$.
\end{theorem}
We can choose $\epr$ to trade-off this storage requirement with the running time requirement of $O\pn{c\epr/\delta}$ -- for example, we can set both the query running-time and per-node storage to $\sqrt{c \gamma\bar{d}/\delta}$ where $\bar{d}=m/n$ is the average degree.
Now for name search $\gamma=2$, and if we choose $\delta= \frac{1}{n}$ and $\alpha=\Theta(1)$, the per-query running time and per-node storage is  $O(\sqrt{m})$.
\begin{proof}
For each set $T$ corresponding to a keyword, and each $t \in T$, we push from nodes $v$ until for each $v$, $r^t[v] < \epr$.  Each time we push from a node $v$, we add an entry to the residual vector of each node $u \in \inneighbors{v}$, so the space cost is $\indegree{v}$.  Each time we push from a node $v$, we increase the estimate $p^t[v]$ by $\alpha r^t[v] \geq \alpha \epr$, and $\sum_{t \in T} p^t[v] \leq \sum_{t \in T} \pi_v[t] = \pi_v[T]$ so $v$ can be pushed from at most $\frac{\pi_v[t]}{\alpha \epr}$ times. Thus the total storage required is 
\begin{equation}
\label{eq:storage}
 \sum_{v \in V}  \indegree{v} (\#\text{ of times $v$ pushed}) \leq  \sum_{v \in V}  \indegree{v} \frac{\pi_v[T]}{\alpha \epr}
\end{equation}

Let $\mathcal{T}$ be the set of all target sets (one target set per keyword).  Then the total storage over all keywords is 
\begin{align*}
\sum_{T \in \mathcal{T}}\sum_{t \in T} \sum_{v \in V}  \indegree{v} \frac{\pi_v[t]}{\alpha \epr} 
& \leq \gamma \sum_{v \in V} \sum_{t \in V}  \indegree{v} \frac{\pi_v[t]}{\alpha \epr} \\
& \leq \gamma \sum_{v \in V}   \indegree{v} \frac{1}{\alpha \epr}
 \leq \gamma \frac{m}{\alpha \epr} .
\end{align*}
\end{proof}

  
\noindent\textbf{Adaptive Maximum Residual}: 
One way to improve the storage requirement is by using larger values of $\epr$ for target sets $T$ with larger global PageRank.  
Intuitively, if $T$ is large, then it's easier for random walks to get close to $T$, so we don't need to push back from $T$ as much as we would for a small $T$.  
We now formalize this scheme, and outline the savings in storage via a heuristic analysis, based on a model of personalized PageRank values introduced by Bahmani et al. \cite{Bahmani2010}.

For a fixed $s$, we assume the values $\pi_s[v]$ for all $v \in V$ approximately follow a power law with exponent $\beta$.
Empirically, this is known to be an accurate model for the Twitter graph -- Bahmani et al.~\cite{Bahmani2010} find that the mean exponent for a user is $\beta = 0.77$ with standard deviation $0.08$.  
To analyze our algorithm, 
we further assume that $\pi_s$ restricted to $T$ also follows a power law, i.e.:
\begin{equation}
  \label{eq:power_law}
   \pi_s[t_i] = \frac{1 - \beta}{\size{T}^{1 - \beta}} i^{-\beta} \pi_s[T].
\end{equation}

Suppose we want an accurate estimate of $\pi_s[t_i]$ for the top-$k$ results within $T$, so we set $\delta_k = \pi_s[t_k]$. 
From Theorem \ref{thm:bidirmain}, the number of walks required is:
$$w = c \frac{\epr}{\delta_k} = c_2 \frac{\epr \size{T}^{1 - \beta} }{\pi_s[T]}$$
where  $c_2 = k^{\beta} c/(1 - \beta)$.
If we fix the number of walks as $w$, then we must set $\epr = w \pi_s[T]/ (c_2 \size{T}^{1-\beta})$. Also, for a uniformly random start node $s$, we have $\EE[\pi_s[T]]=\pi[T]$ (the global PageRank of $T$). This suggests we choose $\epr(T)$ for set $T$ as:
\begin{equation}
	\label{eq:rmax}
\epr(T) = \frac{w \pi[T]}{c_2 \size{T}^{1-\beta}}
\end{equation}

Going back to equation \eqref{eq:storage}, suppose for simplicity that the average $\indegree{v}$ encountered is $\bar{d}$.  
Then the storage required for this keyword is bounded by:
\[ \sum_{v \in V}  \indegree{v} \frac{\pi_v[T]}{\epr} = \bar{d} \frac{n \pi[T]}{\epr} 
= \frac{m c_2 \size{T}^{1-\beta}}{w} . \]
Note that this is independent of $\pi[T]$.  There is still a dependence on $\size{T}$, which is natural since for larger $T$ there are more nodes which make it harder to find the top-$k$.
For $\beta=0.77$ 
, the rate of growth, $\size{T}^{0.23}$ is fairly small, and in particular is sublinear in $\size{T}$.

\noindent\textbf{Dynamic Graphs}: So far we have assumed that the graph and keywords are static, but in practice they change over time.  When a keyword is added to some node $T$, the node's reverse vector $y^t$ needs to be added to the sampling data structure for that keyword.  When an edge is added, the residual values need to be updated.  We leave the extension to dynamic graphs to future work.  

\subsection{Experiments}

We conduct experiments to test the efficiency of these personalized search algorithms as the size of the target set varies.  We use one of the largest publicly available social networks, Twitter-2010 \cite{WebAlgorithmics} with 40 million nodes and 1.5 billion edges.  For various values of $\size{T}$, we select a target set $T$ uniformly among all sets with that size, and compare the running times of the four algorithms we propose in this work, as well as the \texttt{Monte Carlo} algorithm.  We repeat this using 10 random target sets and 10 random sources $s$ per target set, and report the median running time for all algorithms.  We use the same target sets and sources for all algorithms.

\noindent\textbf{Parameter Choices}:
Because all five algorithms have parameters that trade-off running time and accuracy, we choose parameters such that the accuracy is comparable so we can compare running time on a level playing field.  To choose a concrete benchmark, we chose parameters such that the precision@3 of the four algorithms we propose are consistently greater than 90\% for the range of $\size{T}$ we used in experiment. We chose parameters for \mc\ so that our algorithms are consistently more accurate than it, and its precision@3 is greater than 85\%.  In the full version we plot the precision@3 of the algorithms for the parameters we use when comparing running time.  

We used $\delta = \pi_s(t_k)$ where $\pi_s(t_k)$ is estimated using Eqn. \ref{eq:power_law}, using $k=3$, power law exponent $\beta=0.77$ (the mean value found empirically on Twitter), and assuming $\pi_s(T) = \frac{\size{T}}{n}$ (the expected value of $\pi_s(T)$ since $T$ is chosen uniformly at random).  Then we use Equation \ref{eq:rmax} to set $\epr$, using $c=20$ and two values of $w$, 10,000 and 100,000. We used the same value of $\epr$ for \bipprprecomp{}, \bipprgrouped{}, and \bipprsampling{}.  For \mc, we sampled $\frac{40}{\delta}$ walks\footnote{Note that \mc\ was too slow to finish in a reasonable amount of time, so we measured the average time required to take 10 million walks, then multiplied by the number of walks needed.  When measuring precision, we simulated the target weights \mc\ would generate, by sampling $t_i$ with probability $\pi_s(t_i)$; this produces exactly the same distribution of weights as \mc\ would.}.  

\begin{figure*}
\centering
\subfigure[Running Time, More Precomputation]{
\label{fig:search_runtime10K}
\includegraphics[width=\columnwidth]{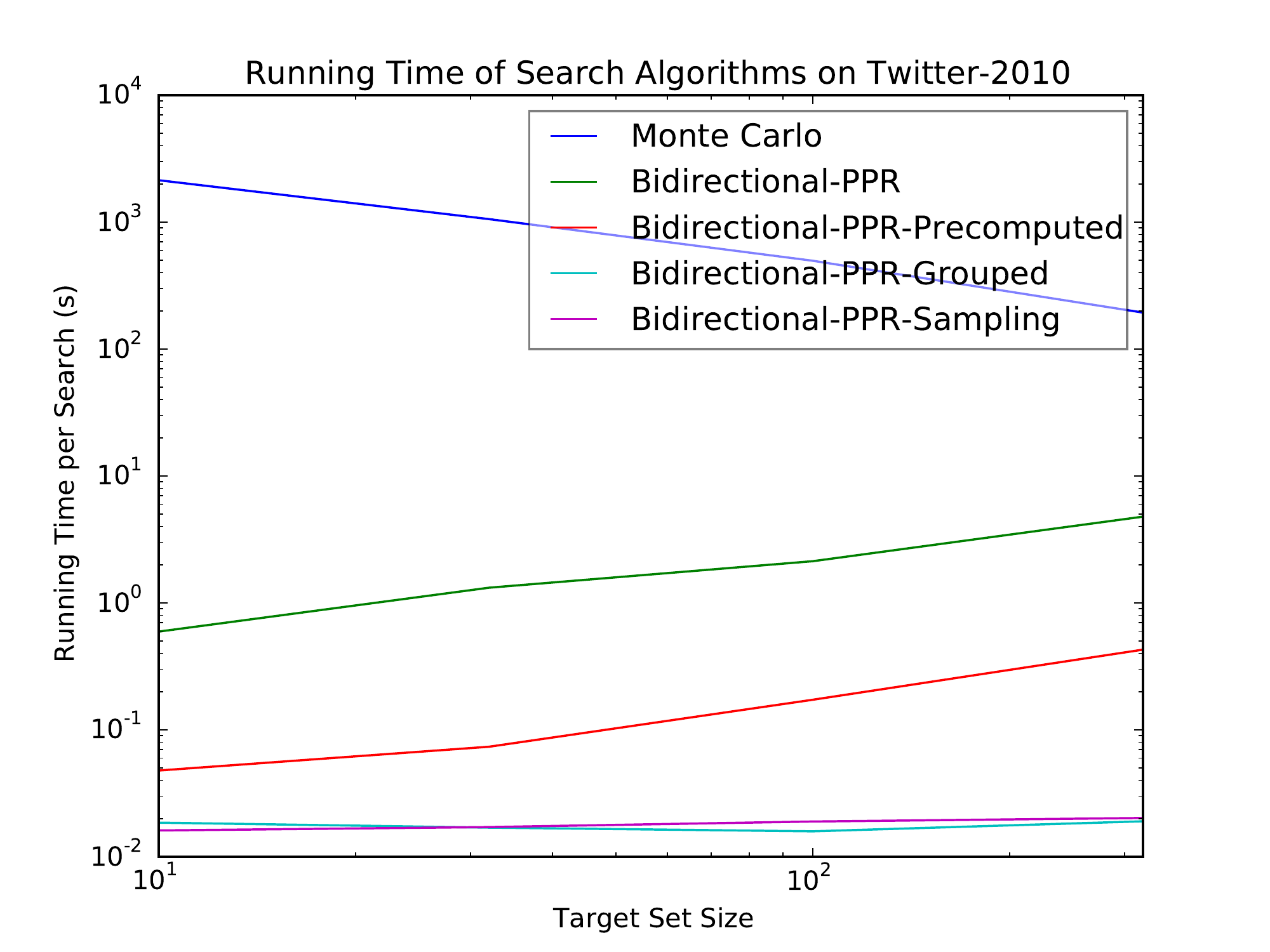}
}
\hfill
\subfigure[Running Time, Less Precomputation]{
\label{fig:search_runtime100K}
\includegraphics[width=\columnwidth]{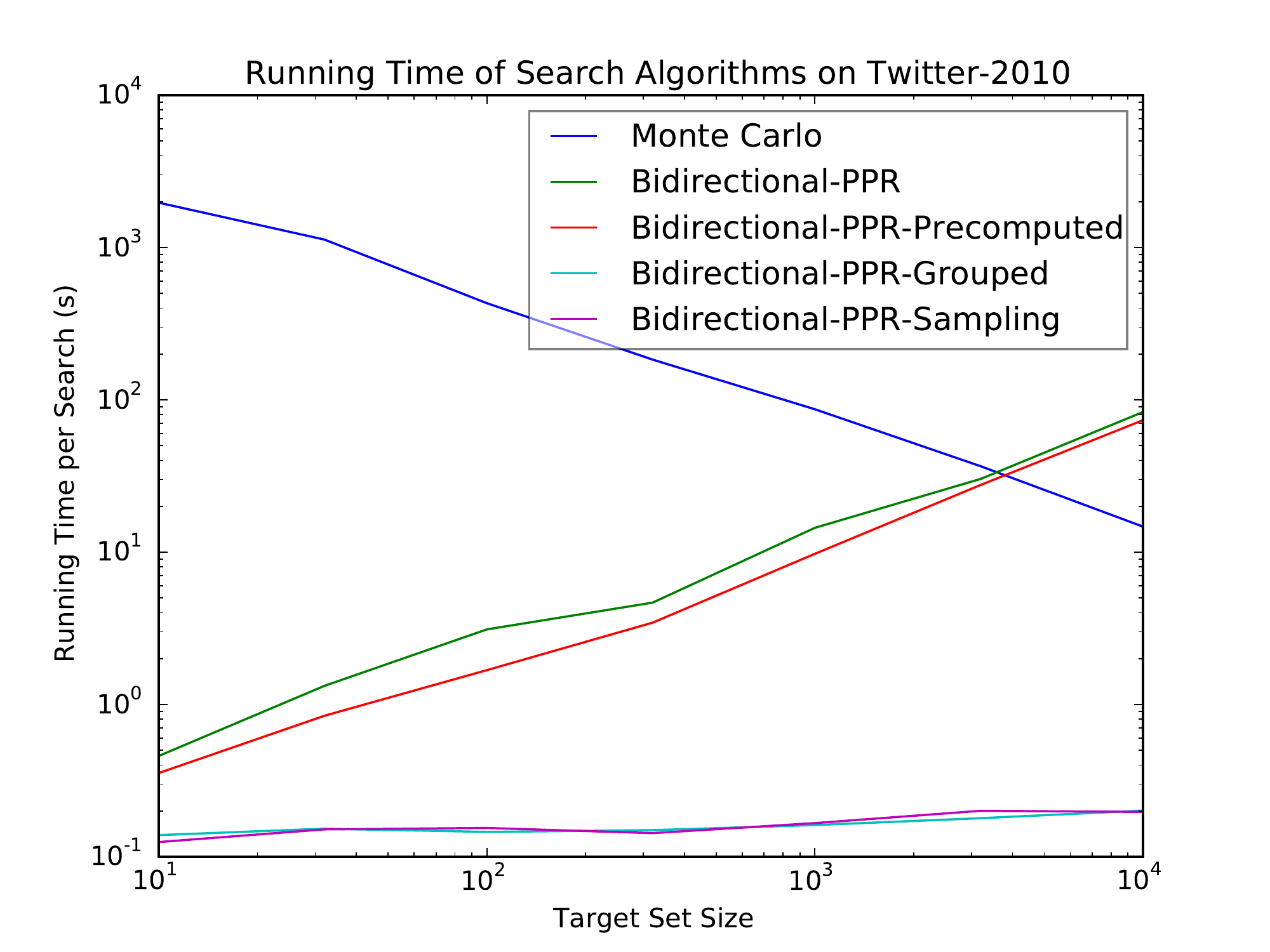}
}
\caption[Comparing PPR-Search Algorithms]{
Running time on Twitter-2010 (1.5 billion edges) on log-scale, with parameters chosen such that the Precision@3 of our algorithms exceeds $90\%$ and exceeds the precision@3 of \mc.\ The two plots demonstrate the storage-runtime tradeoff: Figure \ref{fig:search_runtime10K} (which performs $10K$ walks at runtime) uses more pre-computation and storage compared to Figure \ref{fig:search_runtime100K} (with $100K$ walks).}
\label{fig:search_runtime}
\end{figure*}

\noindent\textbf{Results}:
Figure \ref{fig:search_runtime} shows the running time of the five algorithms as $\size{T}$ varies for two different parameter settings in the trade-off between running time and precomputed storage requirement.   Notice that \mc\ is very slow on this large graph for small target set sizes, but gets faster as the size of the target set increases.  For example when $\size{T}=10$ Monte Carlo takes half an hour, and even for $\size{T}=1000$ it takes more than a minute.
\bippr{} is fast for small $T$, but slow for larger $T$, taking more than a second when $\size{T} \geq 100$.  In contrast, \bipprgrouped{} and \bipprsampling{} are both fast for all sizes of $T$, taking less than 250 ms when $w=10,000$ and less than 25 ms when $w = 100,000$.  

The improved running time of \bipprgrouped\ and \bipprsampling{}, however, comes at the cost of pre-computation and storage. With these parameter choices, for $w=10,000$ the pre-computation size per target set in our experiments ranged from 8 MB (for $\size{T}=10$) to 200MB (for $\size{T}=1000$) per keyword.  For $w=100,000$, the storage per keyword ranges from 3 MB (for $\size{T}=10$) to 30MB (for $\size{T}=10,000$).

\begin{figure}
\centering
\includegraphics[width=\columnwidth]{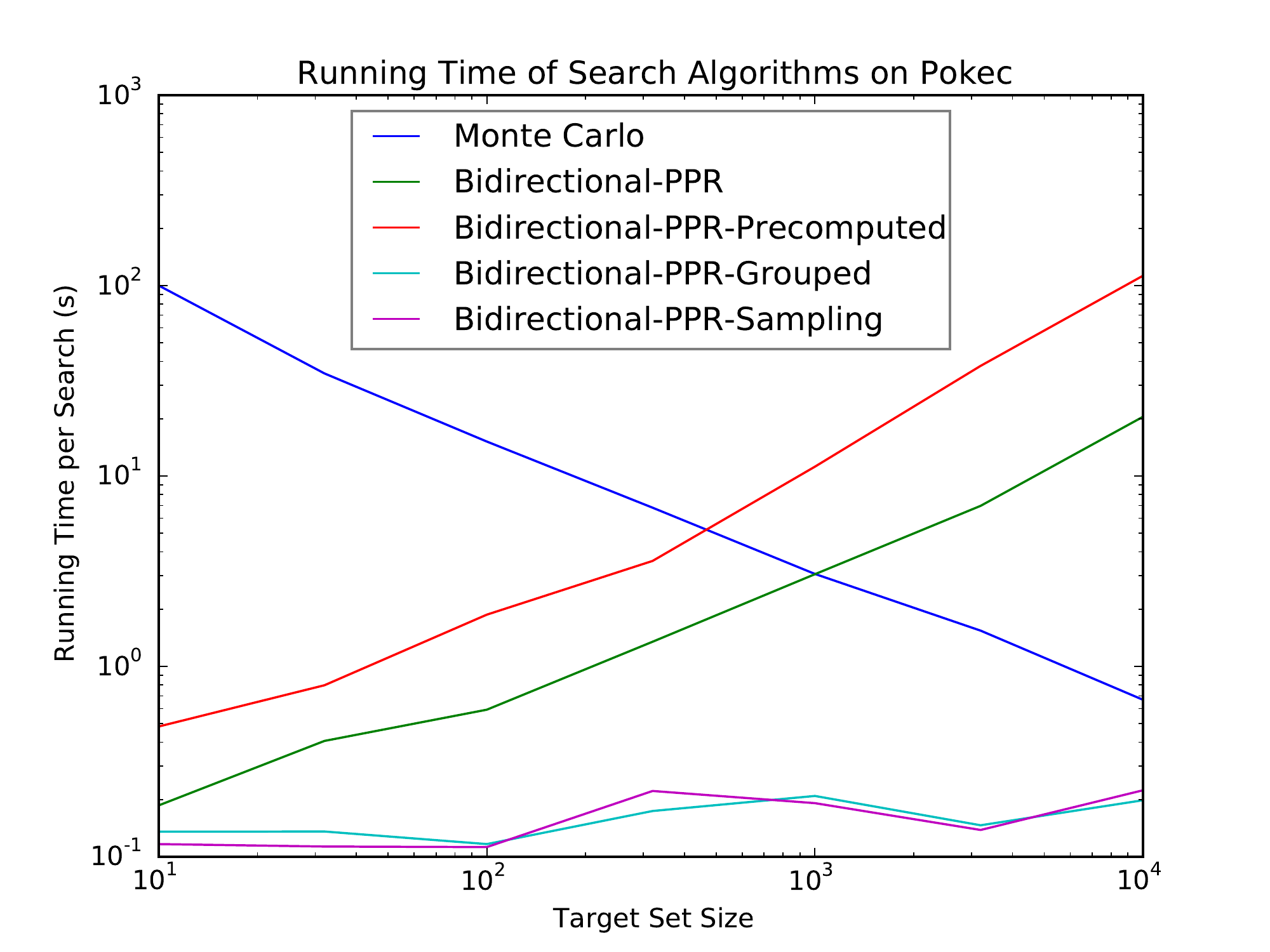}
\caption[Pokec Running Time]{Running time on Pokec (30 million edges) performing 100K walks at runtime.  Notice that \mc\ is slow for small $\size{T}$, \bippr\ is slow for large $\size{T}$, and \bipprgrouped\ and \bipprsampling\ are fast across the entire range of $\size{T}$.}
\label{fig:runtime_pokec}
\end{figure}

To get a larger range of $\size{T}$ relative to $\size{V}$, we also perform experiments on the Pokec graph \cite{SnapProject} which has 1.6 million nodes and 30 million edges. Figure \ref{fig:runtime_pokec} shows the results on Pokec for $w=100,000$.  Here we clearly see the cross-over point where \mc\ becomes more efficient than \bippr, while \bipprgrouped\ and \bipprsampling\ consistently take less than 250 milliseconds.  On Pokec, the storage used ranges from 800KB for $\size{T}=10$ to 3MB for $\size{T}=10,000$.

We implement our algorithms in Scala and report running times for Scala, but in preliminary experiments \bipprgrouped{} is 3x faster when re-implemented in C++, we expect the running time would improve comparably for all five algorithms.  Also, we ran each experiment on a single thread, but the algorithms parallelize naturally, so the latency could be improved by a multi-threaded implementation.
We ran our experiments on a machine with a 3.33 GHz 12-core Intel Xeon X5680 processor, 12MB cache, and 192 GB of 1066 MHz Registered ECC DDR3 RAM.  
We measured the running time of the tread running each experiment to exclude garbage collector time.  
We loaded the graph used into memory and completed any pre-computation in RAM before measuring the running time of the algorithms.


\section{Related Work}

\noindent\textbf{Prior work on PPR Estimation}
The \texttt{Bidirectional-PPR} algorithm introduced in the first half of this work builds on the \texttt{FAST-PPR} algorithm presented in \cite{fastppr} -- for details of prior work on Personalized PageRank estimation, see the section on existing approaches in \cite{fastppr}. 
Although \texttt{FAST-PPR} was the first algorithm for PPR estimation with sublinear running-time guarantees, it has several drawbacks which are improved upon by our new \texttt{Bidirectional-PPR} algorithm:
  \begin{itemize}[nosep,leftmargin=*]
  \item \texttt{Bidirectional-PPR} has a simple linear structure which enables searching; Eqn. \ref{eq:dotprod} shows that the estimates are a dot-produce between a forward vector $x^s$ and a reverse vector $y^t$.  In contrast, estimates in \cite{fastppr} require monitoring each walk to detect collisions with a ``frontier'' set.
  \item \texttt{Bidirectional-PPR} is 3x-8x faster than \texttt{FAST-PPR} for the same accuracy in experiments on diverse networks.
  \item \texttt{Bidirectional-PPR} is cleaner and more elegant, leading to simpler correctness proofs and performance analysis. 
  This also makes it easier to generalize to arbitrary Markov Chains, as done in \cite{generalized_fastppr}.
  \end{itemize}

\noindent\textbf{Comparison to Partitioned Multi-Indexing}
For personalized search, our indexing scheme is partially inspired by the Partitioned Multi-Indexing (PMI) scheme of Bahmani et al. \cite{Bahmani2012}.  
Similar to our methods, PMI uses a bidirectional approach to rank search results according to shortest path distance from the searching user.
Shortest path is easier to estimate than PPR, due to the fact that shortest path is a metric; moreover, shortest path is believed to be a less effective way of ranking search results than PPR.
From a technical point of view, PMI is based on `sweeping' from closer to more distant targets based on a distance oracle; in contrast, we use sampling to find the most relevant targets.

\noindent\textbf{Prior work on Personalized PageRank Search}
In \cite{berkhin2006bookmark}, Berkhin builds upon the previous work \cite{Jeh2003} and proposes efficient ways to compute the personalized PageRank vector $\pi_s$ at runtime by combining pre-computed PPR vectors in a query-specific way.  In particular, they identify ``hub'' nodes in advance, using heuristics such as global PageRank, and precompute approximate PPR vectors $\hat{\pi}_h$ for each hub node using a local forward-push algorithm called the Bookmark Coloring Algorithm (BCA).  
Chakrabarti \cite{chakrabarti2007dynamic} proposes a variant of this approach, where Monte-Carlo is used to pre-compute the hub vectors $\hat{\pi}_h$ rather than BCA.

Both approaches differ from our work in that they construct complete approximations to $\pi_s$, then pick out entries relevant to the query. 
This requires a high-accuracy estimate for $\pi_s$ even though only a few entries are important.  
In contrast, our bidirectional approach allows us compute only the entries $\pi_s(t_i)$ relevant to the query.  



\section{Acknowledgments}
Research supported by the DARPA GRAPHS program via grant
FA9550-12-1-0411, and by NSF grant 1447697.  One author was supported by an NPSC fellowship.

\bibliographystyle{abbrv}
\bibliography{PPR-Search-refs,PPR-refs}

\balancecolumns

\pagebreak
\appendix

\section{More Experiment Plots}
\begin{figure}[t]
\centering

\includegraphics[width=1\columnwidth]{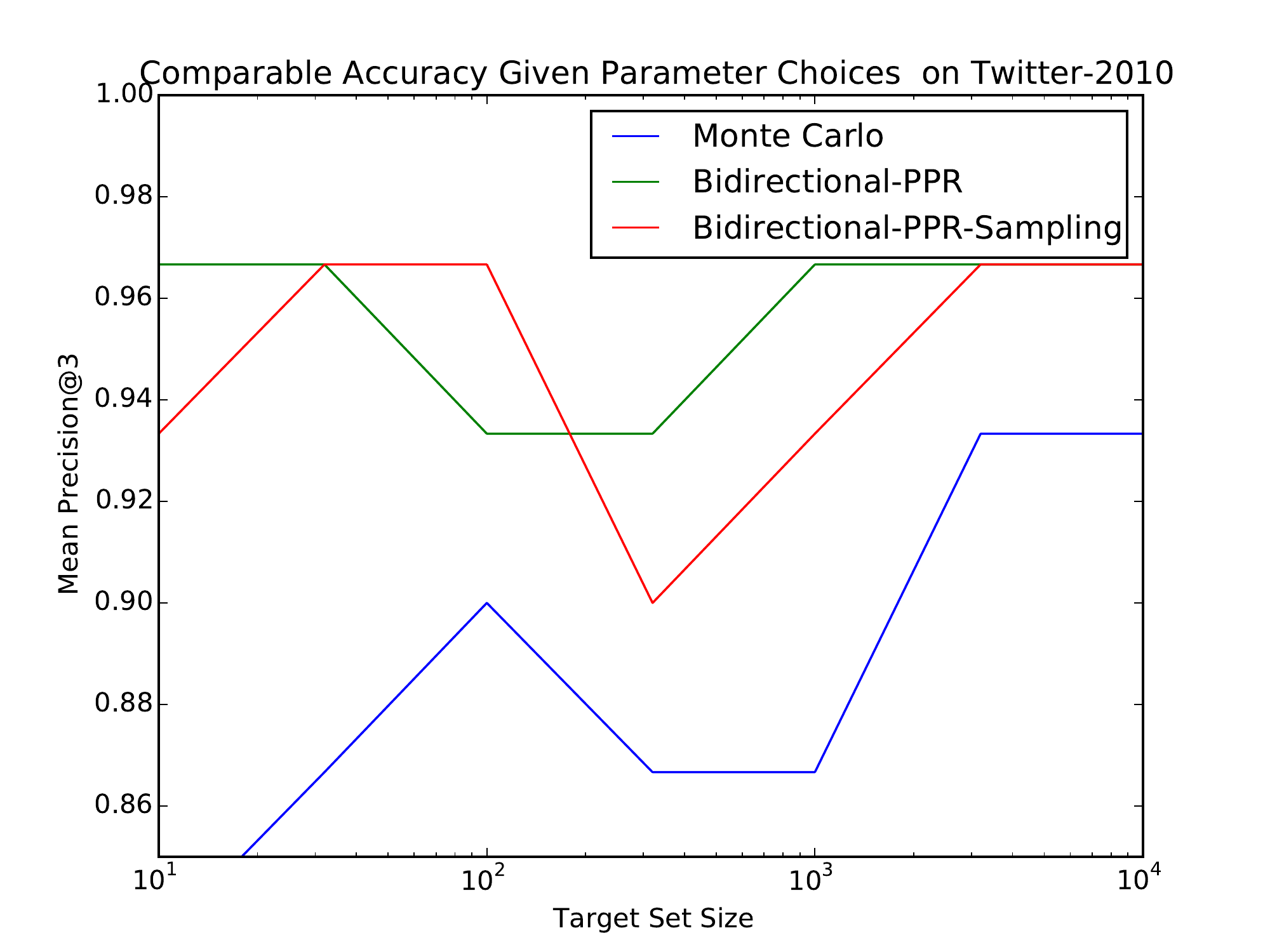}
\caption[Accuracy]{Median precision@3 for the search algorithms we compare.  Notice that the Precision@3 of our algorithms exceeds $90\%$ and exceeds the precision@3 of \mc.}
\label{fig:search_accuracy}
\end{figure}

In Figure \ref{fig:search_accuracy}, we plot the Precision@3 for several search algorithms on Twitter-2010 using the same paramters as the experiments that used $w=100,000$.   Note that \bipprprecomp\ and \bipprgrouped\ compute the same estimates, and these estimates are similar to those of \bippr, so we plot a single line for their accuracy.

\end{document}